\input{aipcheck}

\documentclass[
    ,final            
  ]
  {aipproc}

\layoutstyle{6x9}
\makeatletter
\def\lsim{\compoundrel<\over\sim}
\def\compoundrel#1\over#2{\mathpalette\compoundreL{{#1}\over{#2}}}
\def\compoundreL#1#2{\compoundREL#1#2}
\def\compoundREL#1#2\over#3{\mathrel
      {\vcenter{\hbox{$\m@th\buildrel{#1#2}\over{#1#3}$}}}}
\makeatother

\begin{document}
{\begin{flushright}
 {      CERN-PH-TH/2009-193}
\end{flushright}}
\vfill

\title{Beyond the Planck scale}

\classification{04.60.-m, 04.70.Dy, 11.25.-w, 11.55.-m}
\keywords      {Quantum gravity, black holes, cosmology}

\author{Steven B. Giddings}{
  address={Department of Physics, University of California, Santa Barbara, CA 93106;},  altaddress={PH-TH, CERN, 
1211 Geneve 23, Switzerland}
}



\begin{abstract}
 I outline motivations for believing that important quantum gravity effects lie beyond the Planck scale at both higher energies and longer distances and times.  These motivations arise in part from the study of ultra-high energy scattering, and also from considerations in cosmology.  I briefly summarize some inferences  about such ultra-planckian physics, and clues we might pursue towards the principles of a more fundamental theory addressing the known puzzles and paradoxes of quantum gravity.
\end{abstract}

\maketitle


 While the conference title was ``The Planck Scale,''  a key point that I tried to make is that too sharp a focus on this scale could possibly be misleading, and I believe we need to broaden our focus to make progress on the problem of quantum gravity.   Since I have written extensively on these matters, here I will simply give an overview of some basic points. I have in particular done a very incomplete job of referencing, leaving that to the more complete discussions in the references.

 A possible  analog of our theoretical situation is that preceding the unravelling  of the physics of the atom, and the replacement of classical mechanics by quantum mechanics.  There, viewed purely as a {\it theoretical} problem (in the absence of data about the hydrogen spectrum, {\it etc.}), a classical physicist may have believed that the proper way to resolve the singularity in the classical motion as the electron spirals into the nucleus was to find short distance corrections to classical physics, {\it e.g.} at some scale like the nuclear scale.  But, with benefit of hindsight (and {\it experiment}), we know this is completely wrong; classical mechanics is replaced by the fundamentally new principles of quantum mechanics at a much greater scale, roughly the Bohr radius, $a_0$.  
 
 To motivate our discussion of gravity, let us first note that a complete theory of quantum gravity should describe the physics of ultrahigh energy collisions, with $E\gg M_D$.  (We will consider $D$-dimensional gravity, and $M_D$ denotes the Planck scale.)  The need for a complete theory to address this problem just rests on two very general statements.  {\it Lorentz invariance} of the theory (about a background corresponding to Minkowski space) tells us that we can consider a single particle state that is boosted to arbitrarily high energies and momenta.  Then, a very weak notion of {\it locality} indicates that we can perform such boosts in opposite directions on particles with distant separations (say, light years).  This is all you need to set up an ultrahigh-energy collision.\footnote{While Lorentz invariance violation would change the nature of the problem, it would seem to reappear in a different guise, for example in the physics of black holes, like in the galaxy, formed from states that are not ultrarelativistic; such violation is also phenomenologically problematic.}  Moreover, in recent years we have even been investigating models where, due to large or strongly-warped extra dimensions, the fundamental Planck scale $M_D$ could be as low as $\sim TeV$, and such dynamics could be visible at LHC -- for one review of this exciting possibility, see \cite{Giddings:2007nr}.
 
 In the general ultraplanckian Gedanken experiment, the basic control parameters are the energy, and the impact parameter, $b$ -- after all, these are essential parameters in our high-energy experiments at real colliders.  At $E\gg M_D$, there are good reasons to believe that some important features of the scattering are given by a semiclassical picture.  Classically, for sufficiently small impact parameter, one expects to produce a black hole, plus some radiation as this black hole ``balds.''  Quantum corrections discovered by Hawking tell us that the black hole then evaporates.  So, we expect  an initial state of two high-energy particles, and a final state approximated by Hawking radiation.
 
 The trouble is, this leads to an apparent paradox seemingly driving at the heart of the problem of reconciling quantum mechanics with gravity.\footnote{For some reviews, see \cite{SGinfo,Astrorev}.}  Stephen's calculations have as their modern form what is called the ``nice slice'' argument, which more carefully treats the rough picture in which by locality/causality, there are internal degrees of freedom correlated with the outgoing radiation, and the information associated with these degrees of freedom cannot escape before the black hole evaporates.  This gives a mixed state density matrix $\rho$ describing the final state, with missing information parametrized by the entropy 
 \begin{equation}\label{entropy}
 S= -Tr(\rho\log\rho),
 \end{equation}
  and evolution to this from a pure state implies a violation of quantum mechanics.
 
 However, Banks, Peskin and Susskind\cite{BPS} examined such evolution more carefully, and concluded that through its effects on virtual processes, the natural outcome would be to produce an ambient state behaving like a thermal ensemble at a temperature $T\sim M_D$ -- in painful contradiction with experience!  The other logical possibility, that the information is left behind (or, comes out unitarily after the black hole reaches mass $M\sim M_D$) implies  long-lived remnants, which through the ``infinite-species problem" lead to remnant production instabilities, again in contradiction with every day experience.  (See, e.g., \cite{WABHIP,Susskind} for further discussion.)  So, information can't escape, can't be destroyed, and can't be left behind, and that is the essence of the ``information paradox."
 
This logic just rests on very basic principles: locality/causality, quantum mechanics, and Lorentz invariance (or a local version of it on long distance scales).  Assuming we have not made a silly mistake, we therefore learn one of these needs to be modified.  As I have alluded, modifications of quantum mechanics and Lorentz invariance both appear fraught with difficulties.  But, when you probe it, particularly in a gravitational theory, locality is a remarkably soft concept that is difficult to formulate.  This suggests that it, at least, could be modified in some way.  Moreover, with respect to the semiclassical picture of black hole formation, locality would need to be modified at scales comparable to the Schwarzschild radius of the black hole -- in a high energy collision, this grows with $E$ as
\begin{equation}\label{schrad}
R(E) \propto M_D^{-1} (E/M_D)^{1/(D-3)}\ 
\end{equation}
in units $\hbar=1$.
At ultraplanckian energies, this can be a {\it macroscopic} scale.  While, by arguments of Page\cite{Page} information reemission only has to happen by a time $\tau_i\sim R(E)S(E)$ (where $S(E)\propto R(E)^{D-2}$  is the Bekenstein-Hawking entropy), this represents a major departure from calculations in local quantum field theory in curved space.  Coming back to the analogy of the atom, this suggests that, like there, perhaps some new principles of quantum gravity become important at much longer distances than the previously-expected Planck scale.

There are two major proposals for a complete theory of gravity -- what do they say?  So far, loop quantum gravity is not able to sharply treat the problem of scattering small perturbations about a state that is well-approximated by Minkowski space.  So, let us turn to the question in string theory, which we will consider within the context of a more general discussion of the perturbative approach to such scattering.

\begin{figure}
  \includegraphics[height=.55\textheight]{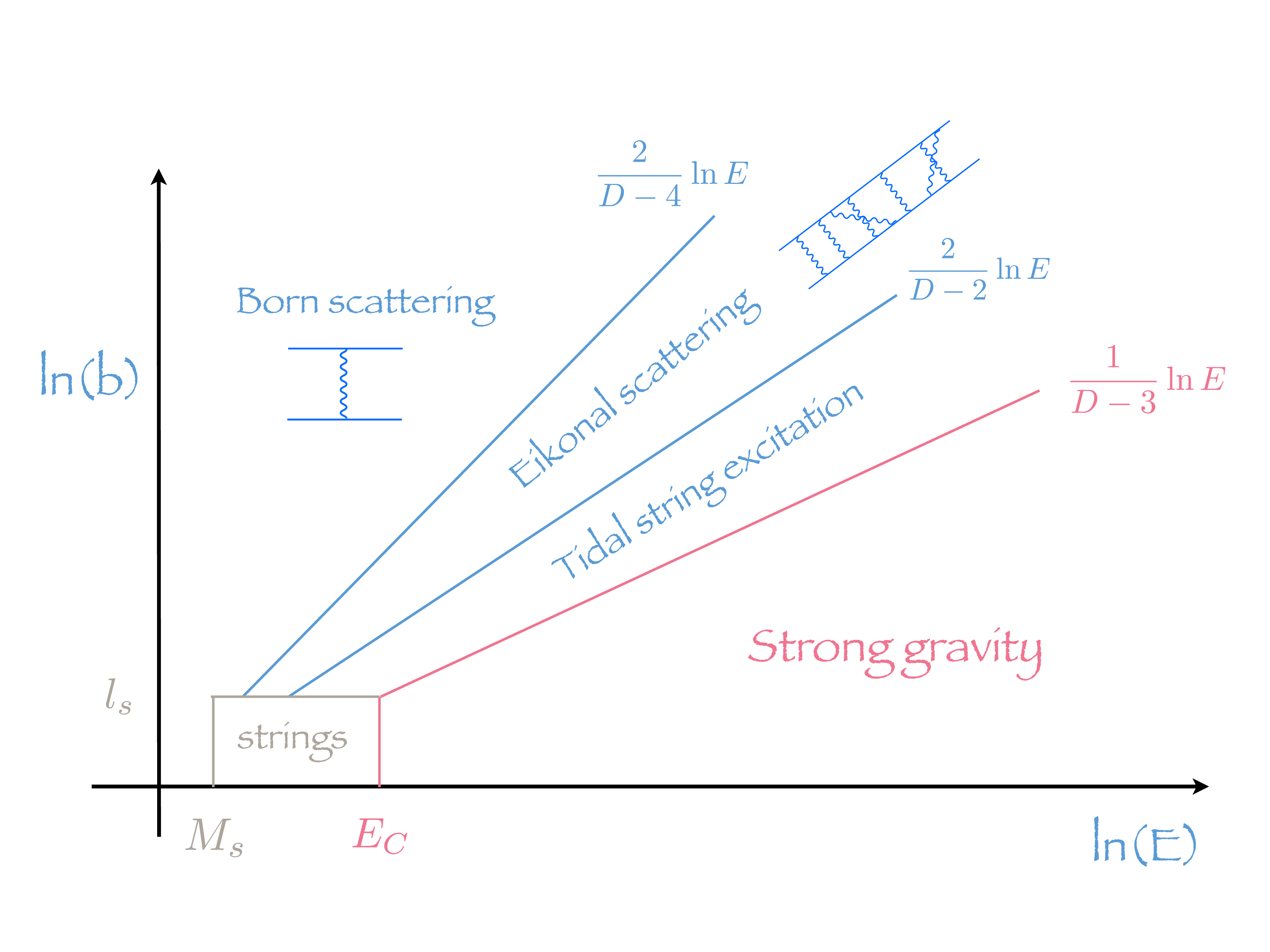}
  \caption{A schematic phase diagram of different high-energy scattering regimes, as a function of impact parameter and energy.  In string theory,  there will be a regime around $E\sim M_{s}= 1/l_s$ where string excitations become relevant  ($E_C$ denotes the black hole/string correspondence point\cite{HoPo}, where $R(E)\sim l_s$), and also a higher-energy regime of ``tidal string excitation." }
\end{figure}

Much of the provisional picture we have is summarized in the ``phase-diagram" of the figure; there is more discussion in the references \cite{LQGST,GGM,GiSr,GiPo} (and important earlier references include \cite{ACVone, ACVtwo, ACVthree,ACVfour}).  For $E\gg M_D$ and  impact parameter decreasing from infinity, which also corresponds to increasing momentum transfer $t=-q^2$, initially scattering is via single graviton exchange.  If the colliding bodies are strings, one na\"\i vely might have guessed that string effects could be important when one reaches impact parameters comparable to the length of a string that could be made with $E$, namely $b\sim M_s^{-2} E$, with $M_s$ the string mass scale.  However, these effects are found to be small.  Instead, what apparently happens is one enters a regime where the interaction is through exchange of multiple single gravitons, described by ladder (or crossed ladder) diagrams, as shown.  These can be summed up to give eikonal amplitudes, which are as usual expected to reflect the physics of the semiclassical approximation.  In particular, with decreasing impact parameter, larger loop order $N$ becomes dominant.  This means that the total momentum transfer $q$ can be divided up into individual momentum transfers of size $k\sim q/N$ for the individual graviton rungs in the ladder.  This division is part of the explanation why exciting oscillatory states of the string does not appear  a central issue -- the individual momentum transfers are too small to do so.  

Given the eikonal/classical relation, one should consider the classical geometry.  This is well-approximated by a pair of colliding Aichelburg-Sexl solutions.  It is true that a string can become excited crossing one such shock wave -- in what one might interpret as a collective effect of the many graviton rungs in the preceding discussion, giving the string a tidal kick.  Indeed, some string theorists had conjectured that such string excitation is a fundamentally important effect in this context, and might even prevent the formation of a black hole.  This was examined, however, in \cite{LQGST,GGM}.  There, by treating the problem of string propagation in such a metric, we argued that while it is true that strings can become excited in a high-energy collision at sufficiently small impact parameter -- just as other composite objects, {\it e.g.} hydrogen atoms, could become tidally excited -- this apparently happens on different time scales than black hole formation.  Specifically, \cite{Penrose,EaGi} showed that in the collision of two Aichelburg-Sexl shockwaves, a trapped surface forms at impact parameters $b\lsim R(E)$.  This surface, hence the black hole, forms {\it before} the shock waves collide, whereas the colliding strings would spread out {\it after} they collide, essentially because of causality.  Thus, one apparently has a picture where a black hole forms with the colliding strings captured within, and at that point we don't know of any string effects that would affect the outcome, by for example allowing the strings to escape.
 
 Thus, in the perturbative quantum theory, what does happen at impact parameters $b\sim R(E)$ appears to be something different.  Namely, while we have described exchange of single gravitons, one has higher diagrams in the perturbative series, corresponding to tree-level diagrams drawn with  gravitons, and then connected to the external high-energy particles.  These are suppressed relative to the iterated single-graviton diagrams by powers of $R(E)/b$ making them relatively unimportant for $b\gg R(E)$.  But, at $b\sim R(E)$, this power series in $R(E)/b$ is apparently badly divergent, and thus, corresponding to classical black hole formation, there is a breakdown of perturbation theory.  Note that this is not something that perturbative renormalizability -- which has been considered a central problem of quantum gravity -- addresses.  That is just the question of whether the terms are finite at each order in this series, a separate matter.  Indeed, this, and the connection with the question of how such physics gives unitary quantum scattering, strongly suggests that the more fundamental issue is not {\it renormalizability} but {\it unitarity}.
 
 There are also proposals for a nonperturbative formulation of string theory, through {\it dualities}, the most simple being the proposed duality\cite{Juan} of string theory in $AdS_5 \times S^5$ to supersymmetric Yang-Mills gauge theory, and it is important to ask what they might say.  However, first note that these proposals in a sense don't directly address the information paradox.  The reason is that in the argument for missing information, one needs to describe the local degrees of freedom both inside and outside the black hole, and to describe their correlations; this requires basic notions of localization as formulated for example in terms of local observables, and we don't have a description of such objects in string theory.  However, we can take the ``outside" viewpoint, and ask whether one can for example calculate the S-matrix using the dual formulation.  Such a calculation would be viewed as a great success, and is bound to shed some light on deeper questions.  
 
Extending previous discussions\cite{Polch,Suss,FSS}, there has been a sharper focus on this problem in the past year.  One can first simply ask the question\cite{FSS} whether the familiar form of the S-matrix in the Born, or single graviton exchange, approximation can be extracted from a boundary dual theory, in a limit where the AdS radius $R$ is taken to be large, so that there is a clean identification with scattering in a flat background.  By ``cheating'' a little (that is, assuming certain information about the theory in the bulk of AdS), we showed\cite{GGP} how basic features of the Born amplitude -- the momentum- conserving bulk delta function, and the reduced transition matrix element, $T\sim M_D^{2-D} s^2/t$, can be encoded in the correlation functions of a boundary theory.  This required the appearance of a certain singularity in these boundary correlators, whose coefficient gives $T$, and checks out in fine detail.  This thus gives necessary conditions for boundary correlators  to satisfy in order to reproduce bulk physics.  
Moreover, \cite{HPPS} conjectured that in boundary conformal theories with appropriate behavior in a large-$N$ limit, correspondingly roughly to excited string states growing heavy as compared to the effective mass $1/R$ arising from the curvature, one would find such singularities, and \cite{HPPS} provided some evidence for this.  

If it can be shown that some class of true boundary theories indeed satisfies these non-trivial necessary conditions, that would be very interesting.  However, there is another issue described in a subsequent paper \cite{GaGiII}.  If one really imagines trying to use the boundary theory to define the bulk theory, and so only has access to incoming data corresponding to sources in the boundary CFT, one must show that these sources can be chosen to reproduce the kinds of bulk wavefunctions required to yield a well-defined scattering theory.  The various constraints on the boundary wavepackets serve as an obstacle to this -- basically the corresponding bulk scattering states have tails that die as a power of distance, and do not clearly give a basis of states that are localized enough to probe the fine-grained structure of the S-matrix.  The success outlined in the preceding paragraph depended on the small ``cheat'' of assuming that bulk information could be used to separate out part of the problematic amplitudes, and it is not clear how this can be done if one truly had access {\it only} to the boundary theory.  In short, then, for more than one reason it remains an apparently nontrivial question whether or not such a dual formulation of string theory serves as a non-perturbative definition that is sharp enough to extract a detailed unitary S-matrix and address the question of its properties in the strong-gravity regime.  

In the absence of detailed predictions from existing proposed approaches to quantum gravity, we can ask whether we can see outlines of the general features 
of such a theory.  First, recall that we see strong indications for important effects at distance scales that grow with energy, as (\ref{schrad}).  Secondly, so far as we've seen, these are associated with a breakdown of the perturbative theory, and its need for a non-perturbative completion -- apparently as opposed to other effects, such as due to extendedness of strings.  Finally, there are good indications that these are associated to a revision of the usual formulations of locality.

There are several reasons to question conventional notions of locality.  First is the information ``paradox;" locality is plausibly the least robust of the underlying assumptions, and thus should perhaps be the one to yield.  As noted, with respect to the semiclassical picture, such effects would have to appear to ``relay'' information over distances of order $R(E)$, a macroscopic distance.  The second is the gravitational growth of scattered objects with energy; this is evidenced for example by the impact parameter at which a test particle scatters at a given fixed angle off a high-energy source growing with energy like $R(E)$, or, in the two-body context, by the perturbative breakdown corresponding to strong gravitational physics at this  radius.  Third, as we will describe more shortly, locality can be probed by local observables; while these don't exist in gravity, there are arguments for the plausible existence of approximately local observables, but such constructs apparently break down in related contexts, due to strong gravitational effects.

If we want to probe the basic physics with the correct properties, we can ask a sequence of increasingly detailed questions.  First, we can ask where its {\it correspondence boundary} lies, that is, where existing local QFT requires modification.  Second, we can ask what mechanism or mechanisms explain the underlying physics.  And third, we can ask what physical and mathematical framework is needed to replace local QFT, and how local QFT emerges from it in an appropriate limit.  

Given what we've described, let us address the first two questions.  There have been various proposals giving the correspondence boundary for such new physics.  Certainly it is commonly believed that local QFT breaks down when curvatures become planckian in size; or those believing in modified dispersion relations and Lorentz violation might suggest a critical energy $\sim M_D$ at which it must be replaced.  In string theory, another proposal is the string uncertainty principle\cite{Vene,Gross}, suggesting an additional term in the position uncertainty, $\Delta x\sim 1/\Delta p+ \Delta p/M_s^2$.  These all basically refer to single-particle properties; another set of proposals are various ``holographic" bounds, essentially suggesting that there is something wrong with local QFT when one considers a state with information greater than the Bekenstein-Hawking entropy, or in the appropriate units, bounding surface area, of a region of space.

After the preceding discussion, one is lead to a different proposal.  In QFT in a weakly-curved background, the physics of multiparticle states is described in terms of Fock space; for example
\begin{equation}
\phi_{x,p} \phi_{y,q}|0\rangle
\end{equation}
gives a state of two particles in minimum-uncertainty wavepackets with approximate positions $x,y$ and momenta $p,q$, where we use creation operators formed by integrating the field operator against the wavepacket.  We expect such a Fock space description to fail when we violate a bound
\begin{equation}
|x-y|^{D-3}> G|p+q|,
\end{equation}
-- the left hand side involves the separation, and the right hand side the center-of-mass momentum, and a constant proportional to Newton's constant.  As suggested in \cite{GiLi,GiddingsSJ,GiddingsBE}, when one violates this {\it locality bound}, gravitational backreaction becomes large and we no longer expect Fock space to be a good description.  Thus, this is a proposal for part of the correspondence boundary; there are generalizations to $N$-particle states\cite{LQGST}, and also in the context of $dS$ space\cite{GiMa}.

Likewise, we seem to have clues about the mechanism; given what we know, one might state that a sort of ``nonlocality principle'' involving a delocalization with respect to the  semiclassical geometry, intrinsic to unitary dynamics of nonperturbative gravity, is at work in such contexts. This may strike one as insufficiently precise, and we would like to say more, but note that it is not vacuous.  Given our discussion, it is in contrast to some other proposals, such as nonlocality due to string (or brane) extendedness, and for the same reason the locality bound is different from the string uncertainty principle.

How should we make progress towards unearthing more basic principles?  Since we are trying to understand how locality can fail, yet be true to an excellent approximation in all known low-energy physics, it is worth investigating how locality is sharply formulated in physics.  We'll discuss observables in a moment, but another way to probe locality is via the high-energy behavior of the S-matrix.  

Independent of one's beliefs about the underlying theory, it is quite plausible that, if it has a solution corresponding to Minkowski space, there is a sharp notion of an S-matrix.\footnote{Or, in $D=4$, an appropriate inclusive generalization summing over soft gravitons.}  This conjecture and further description of expected properties of this object are discussed in \cite{GiSr,GiPo}.

In particular, the $2\rightarrow2$ elements of the S-matrix can be fully parameterized, using a partial wave decomposition, in terms of phase shifts $\delta_l$, and their imaginary parts, the absorptive coefficients $\beta_l$.  The $\delta_l$ can be computed via the Born and eikonal approximations, in the corresponding regimes.  Likewise, the $\beta_l$ can also be calculated, although may have some model dependence.  One contribution to loss of probability in the $2\rightarrow2$ amplitudes is simply soft-graviton Brehmsstrahlung, which was estimated in \cite{GiSr}.  However, when considering the example of string scattering, there can also be absorption in scattering of typical string states, because the strings get excited, and thus contribute to different S-matrix elements; a similar story would hold for other extended objects, {\it e.g.} hydrogen atoms.

The most interesting region, though, is the strong gravity region.  Here basic properties of gravity appear to give us some information about $\beta_l$.  Namely, if a quantum analog of a black hole forms, and then decays into a pure state version of Hawking radiation, the amplitude to get back exactly two particles is of order $\exp\{-S(E)/2\}$.  There is likewise possible information about $\delta_l$ coming from the spectrum of black hole ``resonances\cite{GiPo}.''

An interesting feature is that, both due to this behavior, and even apparently from the eikonal amplitudes, one appears to have amplitudes behaving nonpolynomially  in the momentum at large momentum.  In general studies of the S-matrix, locality is postulated through such polynomiality.  So, there appear to be hints about gravity's lack of the usual locality, both in this, and also in the related lack of a mass gap in the theory.  This behavior, which is worthy of more exploration, is discussed further in \cite{GiPo}.

The S-matrix involves taking the asymptotic viewpoint, but it is also important to consider the viewpoint of observers ``inside'' the system -- {\it e.g.} inside a cosmology or black hole.  Indeed, we are the former, and understanding the physics of the latter is needed to more sharply address the information ``paradox.''  Also, here one makes contact with the other way we have of formulating locality in quantum field theory, as commutativity of gauge invariant local observables outside the light cone.  The problem is that gravity does not admit gauge invariant local observables, since diffeomorphisms translate points, and this raises the puzzle of how we even address this class of questions.  I believe that the correct approach is relational, following ideas going back to Leibniz and Einstein, and also explored by various people in the modern quantum gravity literature.  Specifically, in \cite{GMH}, we studied constructs of diffeomorphism invariant observables, called {\it proto-local observables}, that approximately reduce to local observables.  The reduction only occurs in certain backgrounds; written in an effective field theory approximation, the observables would take the form
\begin{equation}
{\cal O} = \int d^D x \sqrt{-g} B(x) O(x)
\end{equation}
where $O(x)$ is the local observable one expects to recover, and $B(x)$ is an operator that effectively localizes it, by for example having a sharply-peaked expectation value in appropriate states of the theory.  

An important point is that if this is how local constructs are recovered, there are intrinsic limitations to such locality\cite{GMH}.  One might ask if these or related limitations could in fact stand in the way of a sharp derivation of the loss of information in black hole evaporation, resolving the ``paradox.''  This question was examined in \cite{QBHB}.  In short, a sharp derivation of the missing information (\ref{entropy})  requires the density matrix of the Hawking radiation, and that is computed by tracing out the internal degrees of freedom.  However, sharply calculating the state of the combined internal and external radiation involves specifying a state on a ``nice-slice'' that smoothly cuts through the geometry.  It is difficult to make this a gauge invariant concept.  There are actually two possible effects that interfere with a sharp definition of this state, when working on the long time scales, $\tau_{i}\sim R(M) S(M)$ by which the information needs to be emitted\cite{Page}.  One arises from the backreaction of the background that is part of giving a gauge-invariant definition of the state along the preceding lines; the other arises merely from the fluctuations in the Hawking radiation itself, which can have an important long-time effect on the state.  Ref.~\cite{QBHB} proposed that a corresponding lack of a sharp calculation of the information loss resolves the conflict producing the {\it paradox}.  This still leaves the {\it problem} of understanding the dynamics by which the information escapes. The apparent failure of perturbation theory in computing the nice slice state, and moreover in the high-energy scattering problem, suggests that the nonperturbative dynamics that unitarizes the theory is needed for a complete answer.  It's worth noting that there is also a parallel set of arguments in cosmologies like dS, where the nice slices are the usual spatial slices.  Here, one also finds fluctuations providing an apparent obstacle\cite{QBHB} to sharp specification of the corresponding state at long times $\tau\sim R_{dS}S_{dS}$.  An apparently related argument\cite{Nimaetal} shows that if one instead regulates inflation by its termination through slow-roll, there is a similar limit to the duration of inflation for which such regulation works -- after this, eternal inflation is a manifestation of strong fluctuations.

Going further, one expects investigation of how to approximately reproduce local observables to be generally important in cosmology; initial treatment of this include \cite{GMH,GaGi,GiMa} as well as work in progress.  It appears that protolocal observables can be part of the story of the proper handling of the constraints in dS backgrounds, and the question of linearization instability.  A proposed generalization of the preceding limitations on a local QFT description is that such a description won't be valid over a spacetime region larger than $\sim R_{dS}^D \exp\{S_{dS}\}$.  A formulation of the appropriate observables of the theory, and limitations on recovering local constructs, seems an important key to the more general framework and its interpretation.

To summarize the discussion, we should be probing the limits of a local quantum field theory description of nature, and there are good reasons to believe that these limitations extend to distances much larger than the Planck length, $l_{P}$, in certain circumstances.  Given both the information ``paradox" and the apparent breakdown of perturbation theory at large distance scales in high-energy collisions, unitarity is plausibly restored at the price of locality, in the ultimate non-perturbative formulation of the theory. 

We should ask how we can make concrete progress, or gain further hints.  I have argued that the conceptual adjustment we need may be comparable to that of the transition from classical to quantum mechanics, which was guided by experiment -- so our task may not be simple.  But, I think we have some guides.  First, we can investigate properties of the gravitational S-matrix, in accord with general principles of unitarity and analyticity, and taking into account known and expected features of gravity.  Second, we can more deeply investigate the question of how to formulate observables that allow us to recover the usual constructs of local quantum field theory, at least in an approximation.  Finally, if the theory is indeed quantum-mechanical, as I am assuming, we should understand how to give a sufficiently general framework of quantum mechanics to incorporate the physics of gravity.  For example, Hartle\cite{Hartone,Harttwo,HartLH,HartPuri} has formulated a ``generalized quantum mechanics" that goes in the right direction.  But, I believe this still could be too closely tied to spacetime concepts and the notion of a classical ``history," and have proposed that we need to generalize somewhat further.  An initial foray in this direction appears in \cite{UQM}.

Within this or some other quantum framework, our task is to find a theory that is {\it consistent}, but lacks some of the locality properties of local quantum field theory, while recovering them in an excellent approximation at low energies, non-cosmological times/distances, {\it etc.}, where we have never encountered their violation.  There is an essential tension here between non-locality and near-locality -- typical attempts to introduce non-locality into physics lead to massive non-locality, and worse, acausality/inconsistency.  Our theory needs a  certain ``locality without locality.''  Thus, we find a very nontrivial set of constraints to satisfy; but this level of stringency could in turn help guide us on the right path.  I've provisionally referred to the expected outcome as ``non-local (but nearly-local) mechanics,'' recognizing these features as well as the importance of aspects of quantum mechanics, and my hope is that pursuing these and related considerations will bring us closer to understanding a corresponding set of basic principles and mathematical formulation of that mechanics.

\begin{theacknowledgments}
I wish to thank my collaborators, particularly D. Eardley, M. Gary, D. Gross, J. Hartle, M. Lippert, A. Maharana, D. Marolf, J. Penedones, R. Porto, and M. Srednicki, for valuable discussions.  I greatly appreciate the stimulating hospitality of the CERN theory group during the time this was prepared.
This work  was supported in part by the Department of Energy under Contract DE-FG02-91ER40618,  and by grant  RFPI-06-18 from the Foundational Questions Institute (fqxi.org).  
\end{theacknowledgments}



\bibliographystyle{aipproc}   


\IfFileExists{\jobname.bbl}{}
 {\typeout{}
  \typeout{******************************************}
  \typeout{** Please run "bibtex \jobname" to optain}
  \typeout{** the bibliography and then re-run LaTeX}
  \typeout{** twice to fix the references!}
  \typeout{******************************************}
  \typeout{}
 }

\end{document}